\documentclass[a4paper,10pt,twoside]{cpc-hepnp}

\usepackage{CJK}

\usepackage{upgreek,fancyhdr}
\usepackage{multicol}
\usepackage{graphicx}
\usepackage{booktabs}
\usepackage{amssymb,bm,mathrsfs,bbm,amscd}
\usepackage[tbtags]{amsmath}
\usepackage{lastpage}

\usepackage{color}
\usepackage{subcaption}

\begin{document}
\begin{CJK*}{GB}{gbsn}
\fancyhead[c]{\small Chinese Physics C}



\title{Application of SNiPER framework to BESIII physics analysis\thanks{Supported by Joint Large-Scale Scientific Facility Funds of the NSFC and CAS (U1532258), Program for New Century Excellent Talents in University (NCET-13-0342) and Shandong Natural Science Funds for Distinguished Young Scholar (JQ201402), National Key Basic Research Program of China under Contract (2015CB856700).}}

\author{%
      Xin Xia ()$^{\star;1)}$\email{xin.xia@cern.ch}
\quad  Teng Li ()$^{\star;2)}$\email{liteng@hepg.sdu.edu.cn}
\quad  Xing-Tao Huang ()$^{\star;3)}$\email{huangxt@sdu.edu.cn}
\quad  Xue-Yao Zhang ()$^{\star}$
}
\maketitle

\address{%
$^\star$ Shandong University (SDU), Jinan, Shandong 250100, China\\
}

\begin{abstract}
A fast physics analysis framework has been developed based on SNiPER to process the increasingly large data sample collected by BESIII.
In this framework, a reconstructed event data model with SmartRef is designed to improve the speed of Input/Output operations, and necessary physics analysis tools are migrated from BOSS to SNiPER.
A real physics analysis $e^{+}e^{-} \rightarrow \pi^{+}\pi^{-}J/\psi$ is used to test the new framework, and achieves a factor of 10.3 improvement in Input/Output speed compared to BOSS.
Further tests show that the improvement is mainly attributed to the new reconstructed event data model and the lazy-loading functionality provided by SmartRef.
\end{abstract}



\begin{multicols}{2}

\section{Introduction}

The Beijing Spectrometer III (BESIII)~\cite{2009NIMPA.598....7B} is a detector at the Beijing Electron--Positron Collider II (BEPCII).
The accelerator has two storage rings with a circumference of 224\,m and a crossing angle of 22\,mrad.
Its designed peak luminosity is 1$\times$10$^{33}$\,cm$^{-2}$s$^{-1}$ at a beam energy of 1.89\,GeV~\cite{2009NIMPA.598....7B}.
In April 2016, BEPCII successfully reached this goal.
Assuming $10^{7}$\,s data taking time each year, the BESIII detector is able to collect $\sim$10 billion $J/\psi$, $\sim$300 million $\psi^\prime$, 30 million $D\bar{D}$ or 2 million $D^+_SD^-_S$.
The huge amount of data collected makes it possible to study light hadron spectroscopy in the decay of charmonium states and charmed mesons with unprecedentedly high precision~\cite{2008arXiv0809.1869A}.
Since its first collisions in June 2008, BESIII's data volume has reached $\sim$3\,PB and is increasing at a speed of about 0.5\,PB per year.
The BESIII Offline Software System (BOSS) is the main framework that is currently used in BESIII experiment, and its role is very important in the whole offline data processing and physics analysis workflow.
Based on Gaudi~\cite{2001CoPhC.140...45B}, BOSS provides standard interfaces for common software components which are necessary for data processing and analysis. However, Input/Output (I/O) in BOSS requires data format conversion and full-size data read-in, which uses extra CPU time and imposes restrictions on the processing speed~\cite{proc-disc-2006}.
Consequently, I/O is becoming a bottleneck for BOSS, especially for physics analysis.

The Software for Non-collider Physics Experiments (SNiPER)~\cite{sniper} is a software framework for simulation, reconstruction and analysis in a variety of experiments like the Jiangmen Underground Neutrino Observatory (JUNO)~\cite{2016JPhG...43c0401A} and the Large High Altitude Air Shower Observatory (LHAASO)~\cite{2016arXiv160207600D}.
SNiPER is a light-weight, flexible framework with an event data management system which is designed to manage any type of event data. Consequently, no additional data format conversion is needed.
In order to improve the speed of BESIII physics analysis, SNiPER is being applied to the BESIII experiment. 
In this article, details are given of the redesign of the reconstructed event data model, the migration of BESIII physics analysis tools into SNiPER, and a comparison of its performance with that of BOSS.

\section{Redesign of the Reconstructed Event Data Model}

For physics analysis in BOSS, the whole information of each reconstructed event is read in from the Data Summary Tape (DST) files, where data are stored as ROOT trees.
The event data are then converted to Gaudi's data format to be managed by the Gaudi event data store.
This conversion requires extra CPU time and memory, and thus slows down the physics analysis.
In the physics analysis, the input information of each event is processed with analysis algorithms for further event selection.
This kind of full-size information read-in of each event also degrades the speed of analysis.

In SNiPER, the event data management system is designed to use the ROOT format all the way through the whole process.
Event information that is read in from the DST files can be handled directly by  SNiPER's event data management (EDM) system, so no conversion process is needed.
The I/O system is designed to first read only part of the event information for fast event selection. 
Once it meets all selection requirements, the whole event information will be read. 
Therefore, time spent on the data I/O process is significantly reduced, which is similar to the tag-based pre-selection mechanism under the BOSS framework~\cite{tag_boss}.

To use the data management system in SNiPER, a new EDM for the BESIII physics analysis has been designed.
This reconstructed event data model consists of two layers, DstEvent and EvtHeader, as shown in Fig.~\ref{edm}.
The DstEvent contains the full information of physics events, and keeps their original structure in the old DST files.
The EvtHeader is a newly added layer.
With the negligible increase in size (in permillage level), the EvtHeader plays a very important role in the framework.
1) It is the entrance point for the event data service to access new ROOT files.
2) It stores characteristic variables of event, such as the good charged tracks, which can be customized by users to do fast selection without loading the full event information.
3) It stores the SmartRef, which is a smart pointer providing references to events~\cite{smartref}.
The referenced object in DstEvent will not be loaded from input files until it is actually needed.
Therefore, the unnecessary performance overhead can be avoided as a result of the lazy-loading mechanism.

In the analysis algorithm, fast selection is usually applied as soon as the EvtHeader is available.
In fast selection, only the variables in EvtHeader are used without loading the full event into memory.
After fast selection, the full information of survived events will be requested and loaded into memory by the SmartRef for further analysis.
Using this mechanism, significantly less information needs to be read in from disk, which leads to less time consumption in I/O operations.
The stricter the selection is, the faster the I/O operations will be.
These are the main strategies to improve the speed of analysis in general.

\begin{center}
    \begin{center}
    \includegraphics[width=0.47\textwidth]{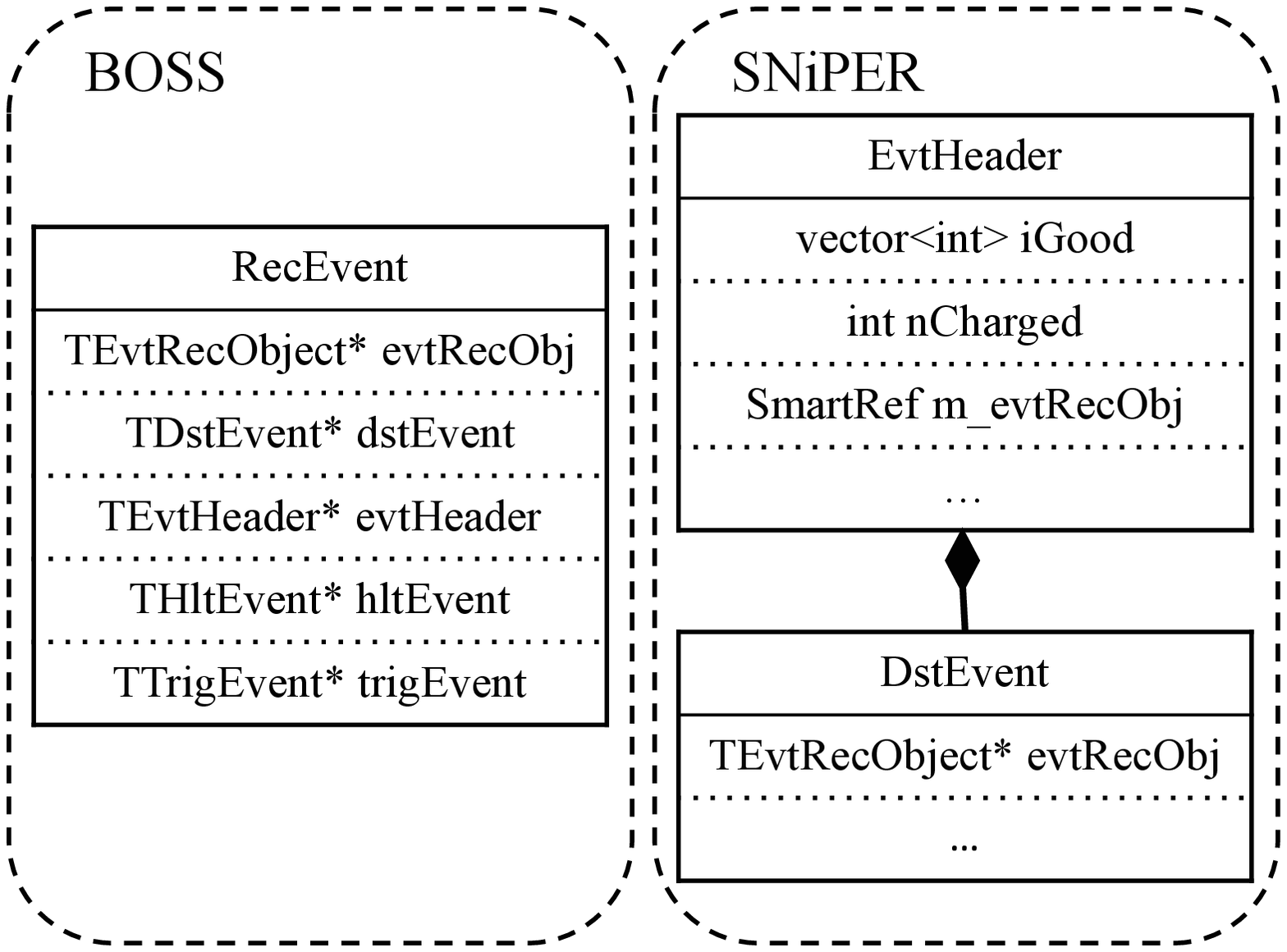}
    \figcaption{\label{edm} Schema of the reconstructed event data model. The left-hand side is the EDM under BOSS, and the right-hand side is the new EDM under SNiPER.}
    \end{center}
\end{center}

\section{Migration of Analysis Tools}

In BESIII physics analysis, several tools are indispensable, including particle identification (ParticleID), vertex fit (VertexFit) and kinematic fit (KinematicFit).

To organize the information left by the final state particles in detectors, a simplified version of the  EvtRecTrack class was imported to integrate all the tracks in the sub-detectors into one logical track from the inside, with their corresponding track IDs stored in the TEvtRecTrack.
To determine the vertex, the primary vertex information and magnetic field information are needed. So the DatabaseSvc and MagneticField were also migrated.
In all these migrated packages, an interface to access the data with the new format was developed. For simplicity, TObject classes are used to substitute the EventObject classes.
The final workflow model used  in SNiPER can be seen in Fig.~\ref{snpscheme}.

\begin{center}
\includegraphics[width=0.48\textwidth]{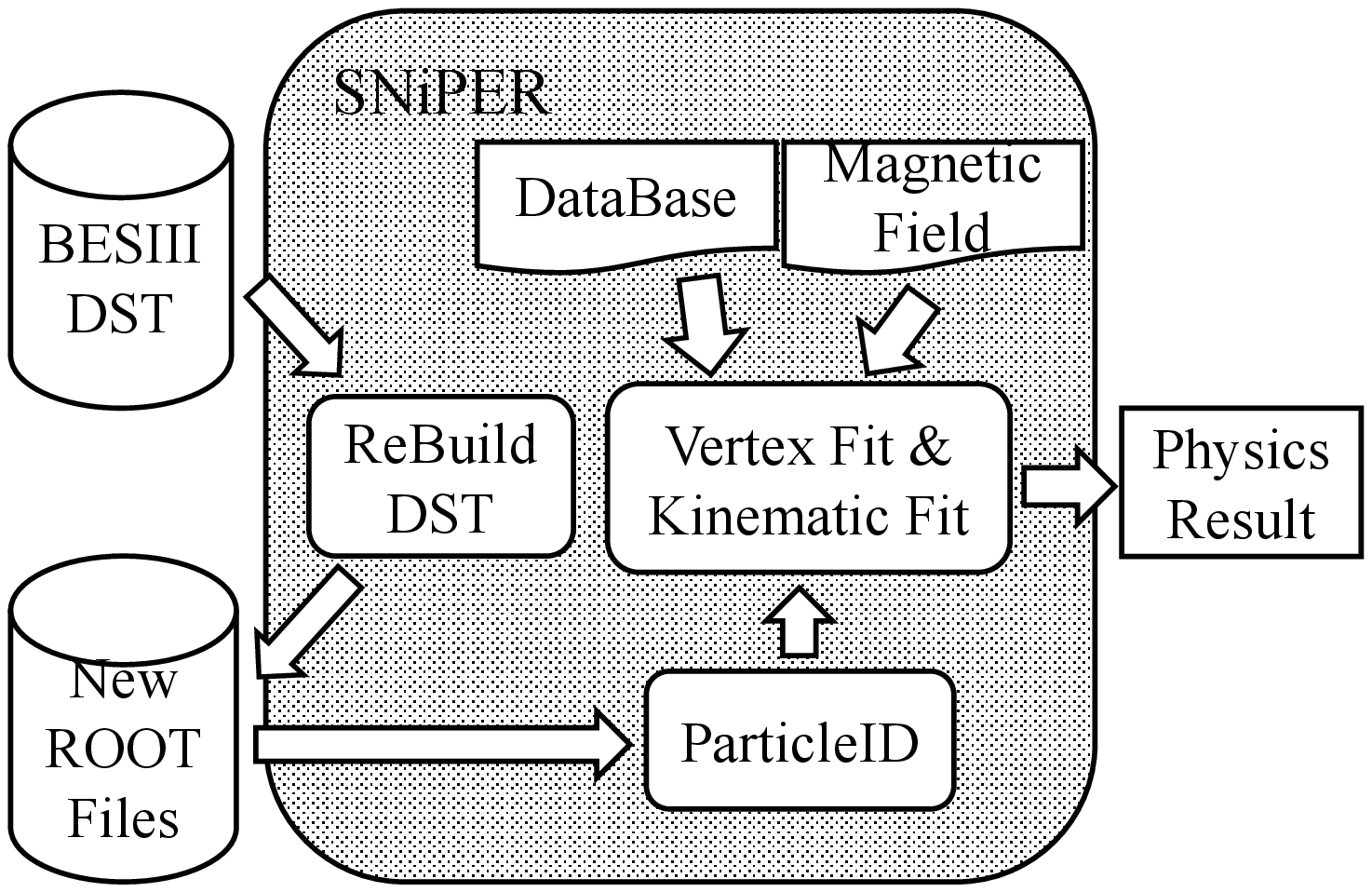}
\figcaption{\label{snpscheme} The new workflow for BESIII analysis in SNiPER.}
\end{center}

\section{Performance and Tests}

In order to validate the migration, we ran the real physics analysis of the process $e^{+}e^{-} \rightarrow \pi^{+}\pi^{-}J/\psi$ at center-of-mass (CM) energy of $\sqrt{s} = (4.260\pm0.001)$\,GeV~\cite{2013PhRvL.110y2001A} with both BOSS and SNiPER.
In the analysis, the $J/\psi$ candidate was reconstructed with lepton pairs ($e^{+}e^{-}$ or $\mu^+\mu^-$), which results in a final state with four charged tracks.
Therefore, the number of charged tracks was required to be no less than 4, which makes the proportion of surviving events approximately 1/200.
In SNiPER, the number of charged tracks is stored in EvtHeader and defined as a tag for fast selection, so only the selected events are fully loaded into memory for further study, which greatly increases the input speed.
In this test, the version of BOSS was $7.0.0$, and 400 DST files were randomly selected for analysis.
Table~\ref{tabcuts} shows the number of surviving events after a series of selections under the two frameworks.
In SNiPER, the results of event selection were exactly the same as for BOSS.

\begin{center}
\tabcaption{ \label{tabcuts} Number of surviving events passing cuts in BOSS and SNiPER}
\footnotesize
\begin{tabular*}{76mm}{clrr}
\toprule
NO.  & Selection                         & BOSS     &  SNiPER   \\ \midrule
1 & Total entries       & 75159209 &  75159209 \\
2 & Charged tracks      & 389796   &  389796   \\
3 & Good charged tracks & 158595   &  158595   \\
4 & Good photon         & 21959    &  21959    \\
5 & Particle identification  & 20483  &  20483 \\
6 & Kinematic fit            & 3748   &  3748  \\
7 & Save result              & 401    &  401   \\ \bottomrule
\end{tabular*}
\vspace{0mm}
\end{center}

\begin{center}
\begin{figure}[htp]
  \centering
  \begin{subfigure}[htbp]{0.32\textwidth}
    \includegraphics[width=\textwidth]{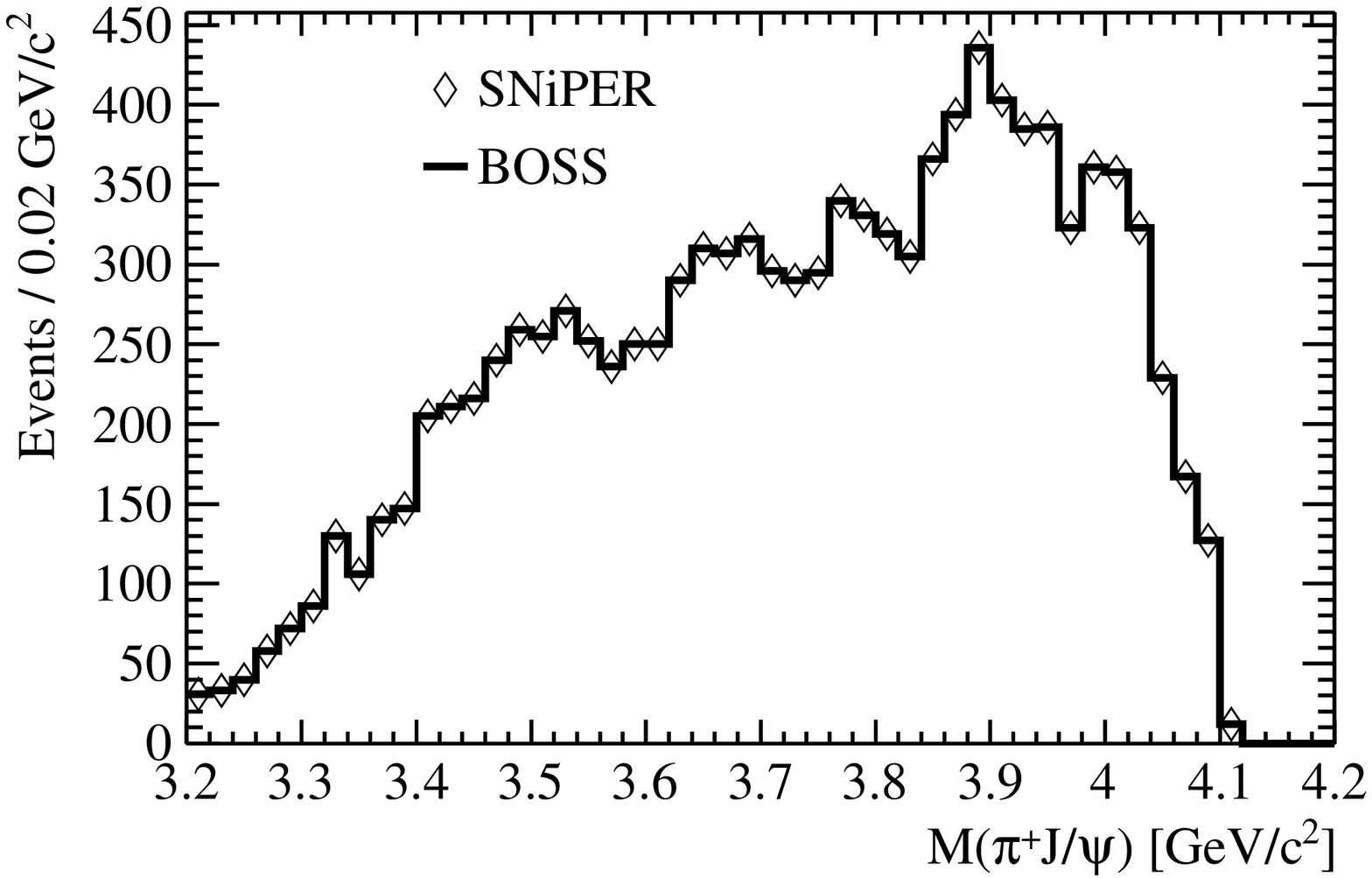}
  \end{subfigure}
  \begin{subfigure}[htbp]{0.32\textwidth}
    \includegraphics[width=\textwidth]{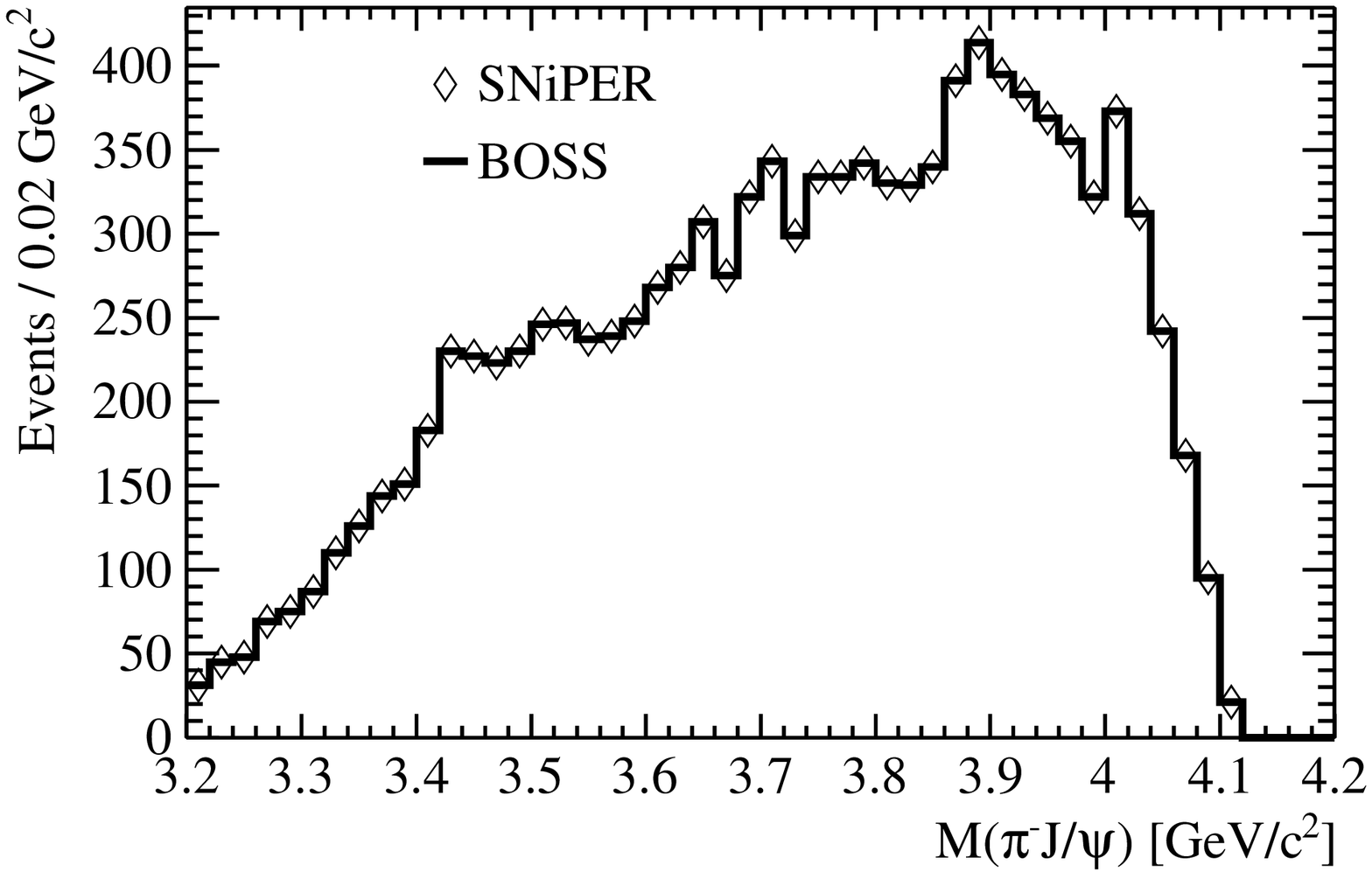}
  \end{subfigure}
  \begin{subfigure}[htbp]{0.32\textwidth}
	   \includegraphics[width=\textwidth]{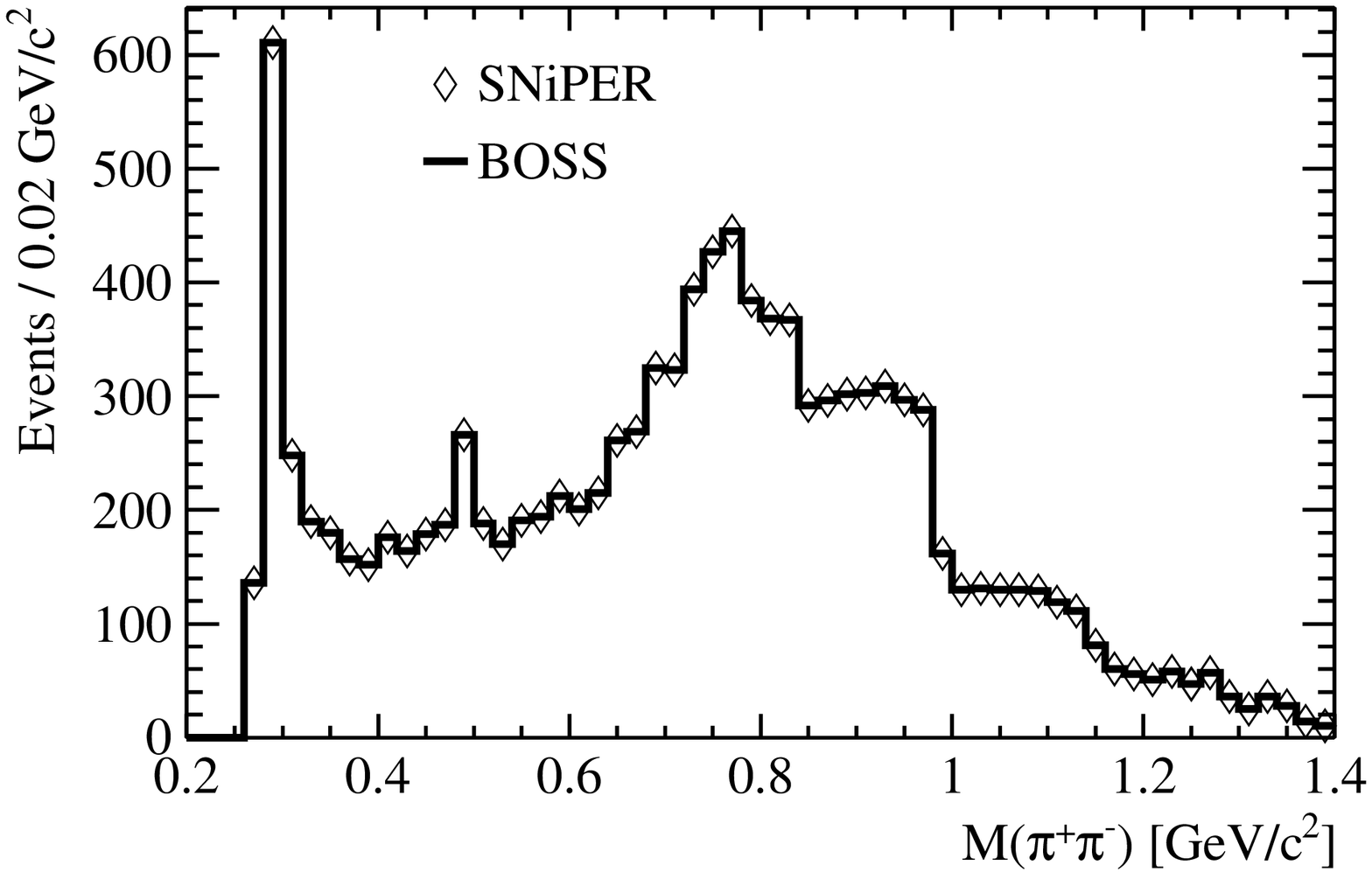}
  \end{subfigure}

  \caption{Comparison of the invariant mass spectra of $M(\pi^+J/\psi)$, $M(\pi^-J/\psi)$, and $M(\pi^+\pi^-)$ in BOSS and SNiPER. The black line represents the results from BOSS, and the open diamonds show the results from SNiPER.}
  \label{imass}
\end{figure}
\end{center}

After checking the step-by-step selections, the invariant mass spectrum of selected $\pi^{+}\pi^{-}J/\psi$ candidates was compared between BOSS and SNiPER using the  whole dataset.
Figure~\ref{imass} shows the distributions of $M(\pi^+J/\psi)$, $M(\pi^-J/\psi)$, and $M(\pi^+\pi^-)$ for the signal events.
The distributions in BOSS and SNiPER agree with each other very well, which means the analysis code migrated to SNiPER works properly.

\end{multicols}
\ruleup
\ruledown

\begin{multicols}{2}

To quantify the improvement gained by the new physics analysis framework, we ran a series of tests with the same analysis of $e^{+}e^{-} \rightarrow \pi^{+}\pi^{-}J/\psi$ at the CM energy of $\sqrt{s} = (4.260\pm0.001)$\,GeV, using the same data sample as the previous test.
Under SNiPER, the total number of charged tracks, which is required to be no less than 4, is added to the EvtHeader as a pre-selection variable.
400 input data files were equally divided into 4 groups, and then submitted to the Portable Batch System in queue \textit{besq}.
The time consumptions were measured under the same hardware environment with a CPU, model  Intel(R) Xeon(R) CPU E5-2680 v3 at frequency of 2.50\,GHz.
The analysis using the old EDM and BOSS consumed $\sim$170.5 minutes on average, while $\sim$44.5 minutes on average were spent with new EDM and SNiPER, which means analysis with the new version is $\sim$3.8 times faster than BOSS.
To investigate where the speed boost comes from, the time consumption of each section was measured and is listed in Table~\ref{timeconsuming}.
The proportion of time consumption for each section under the two frameworks can also be seen in Fig.~\ref{proportion}.

\begin{center}
\tabcaption{ \label{timeconsuming} Comparison of time consumption for BOSS and SNiPER.}
\footnotesize
\begin{tabular*}{77mm}{lccc}
\toprule
            & Framework & EDM \& I/O &  Analysis   \\ \midrule      
BOSS /min   &  1.0      & 135.0      &  34.6       \\            
SNiPER /min &  0.18     & 13.1       &  31.2       \\ \midrule   
Gains       &  5.6      & 10.3       &  1.1        \\ \bottomrule
\end{tabular*}
\vspace{0mm}
\end{center}

These tests indicate that SNiPER itself is running $\sim$5 times faster than BOSS, but the contribution from the new framework is very small due to its fast execution.
With the new EDM with SmartRef, the I/O speed is improved by $\sim$10 times, and it decreases the proportion of I/O time from $\sim$80\% to $\sim$30\%.
The analysis step with SNiPER costs a similar time as BOSS, but the proportion increases significantly due to the improvement of the I/O procedure, which means computing power is concentrated on the real analysis instead of data conversion.

\begin{center}
\includegraphics[width=0.48\textwidth]{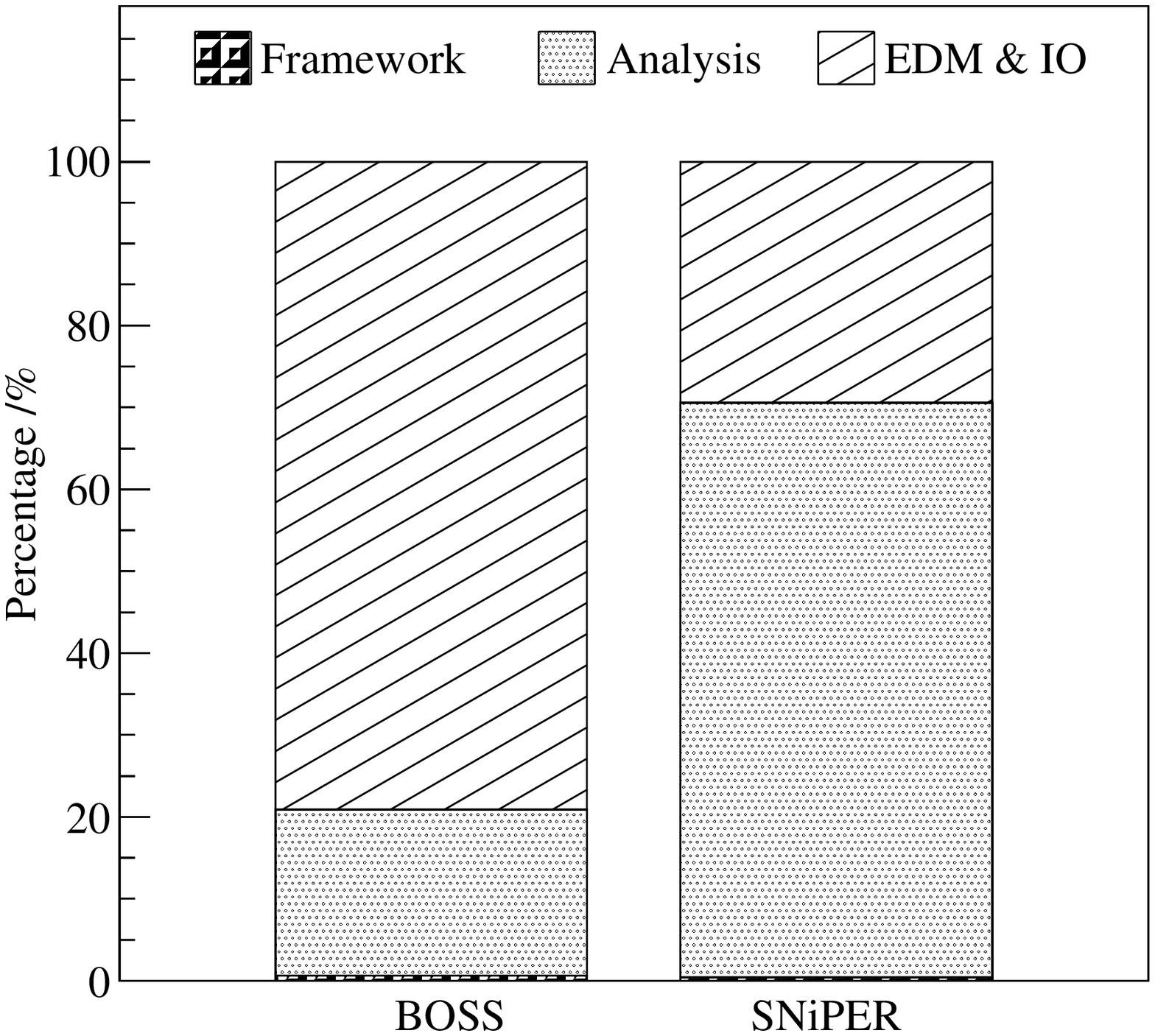}
\figcaption{\label{proportion} Time consumption proportion under BOSS (left) and SNiPER (right) framework.}
\end{center}

\section{Conclusion}

In this article, a new BESIII physics analysis framework based on SNiPER has been introduced, with SmartRef implemented into the reconstructed event data model for fast event pre-selection.
The new framework was tested with a real physics analysis $e^{+}e^{-} \rightarrow \pi^{+}\pi^{-}J/\psi$ at the CM energy of $\sqrt{s} = (4.260\pm0.001)$\,GeV, yielding exactly the same results as the original BOSS framework.
In the test, SNiPER gained 3.8 times improvement in total execution speed, and saved more than 70\% of the time for this specified physics channel.
More tests showed that this improvement is mainly from the new event data model with SmartRef, which gains $\sim$10 times improvement compared to BOSS.

We can conclude that the new physics analysis framework based on SNiPER significantly improves the I/O performance with its redesigned reconstructed event data model using SmartRef.
We can gladly say that this framework is ready for physics analysis in BESIII, and the first stable version of SNiPERMT, which is suitable for the concurrent environment, will be released in 2017.

\vspace{-1mm}
\centerline{\rule{80mm}{0.1pt}}

\end{multicols}

\clearpage

\end{CJK*}
\end{document}